\documentclass[twocolumn,showpacs,pre,floatfix,letter]{revtex4-1}
\ifx\pdfoutput\undefined
\usepackage{graphicx}
\else
\usepackage{epstopdf}
\usepackage{epsfig}
\usepackage{amssymb,amsmath,stmaryrd,tabularx}
\usepackage{wrapfig}
\usepackage{bm}
\fi

\usepackage[center]{subfigure}

\newcommand{\eps}{\epsilon}
\newcommand{\lb}{\left(}
\newcommand{\rb}{\right)}
\newcommand{\be}{\begin{equation}}
\newcommand{\ee}{\end{equation}}
\newcommand{\bea}{\begin{eqnarray}}
\newcommand{\eea}{\end{eqnarray}}
\newcommand{\infinity}{\infty}

\begin{document}
\title{Universal fluctuations and extreme statistics of avalanches near the depinning transition}
\author{Michael LeBlanc$^{1}$, Luiza Angheluta$^{1,2}$, Karin Dahmen$^{1}$ and Nigel Goldenfeld$^{1}$}
\affiliation{
$^1$Department of Physics, University of Illinois at
Urbana-Champaign, Loomis Laboratory of Physics, 1110 West Green
Street, Urbana, Illinois, 61801-3080\\
$^2$Physics of Geological Processes, Department of Physics, University of Oslo, Norway
}


\pacs{64.60.av, 05.40.-a, 05.10.Gg, 61.72.Ff}

\begin{abstract}
We derive exact predictions for universal scaling exponents and scaling functions associated with the statistics of maximum velocities $v_m$ during avalanches described by the mean field theory of the interface depinning transition. In particular, we find a robust power-law regime in the statistics of maximum events that can explain the observed distribution of the peak amplitudes in acoustic emission experiments of crystal plasticity. Our results are expected to be broadly applicable to a broad range of systems in the mean-field interface depinning universality class, ranging from magnets to earthquakes.
\end{abstract}
\maketitle

\section{Introduction}
The depinning transition of a slowly driven elastic interface in systems with quenched disorder is recognized as one of the paradigms of non-equilibrium critical phenomena. As the driving field is slowly increased and fine-tuned to a critical value, the elastic interface undergoes a transition from a pinned state to a moving regime. Near this pinning-depinning transition, the motion of the interface proceeds in spatio-temporal avalanches with scale-invariant statistics. 
Avalanches are observed in a variety of dynamical systems, including the Barkhausen noise in soft magnetic materials~\cite{Zapperi98,Sethna2001}, the charge density wave depinning~\cite{Lee79,Brazovskii04}, the motion of vortices in superconductors~\cite{Field95}, the seismic activity in earthquakes~\cite{Fisher98}, the acoustic emission in mesoscopic crystal plasticity~\cite{miguel2001idf}, and fracture propagation~\cite{Zapperi99}. Although these systems are very different in microscopic detail, the large-scale statistical properties of their avalanche fluctuations appear to be universally captured by the mean field theory of the depinning transition~\cite{Fisher98,Zaiser02,Brazovskii04,Doussal09,Uhl09,Tsekenis11}. 

An avalanche is a burst of forward motion described by a velocity profile $v(x,t),$ where $x$ labels a point along the interface. The motion begins from rest ($v=0$) at  time $t=0$ and ends when the system comes to rest again for the first time, at time $T.$ An elastic interface driven at a slow, constant rate through a disordered medium proceeds forward in a series of avalanches with durations that are distributed according to a power law $P(T)\sim T^{-\alpha}$ in the limit of weak elastic coupling and infinite system size~\cite{Zaiser02}. Another power-law distributed quantity of interest is the global avalanche size, given by $S=\frac{1}{L^d}\int_0^T\int v(x,t)d^dxdt$, which is the total forward motion of the interface center of mass, or, in the domain wall picture, the total change in magnetization during the avalanche~\cite{Doussal09, Kuntz00, Zapperi98}. Here, we will focus on another quantity: the maximum collective velocity during an avalanche $v_m=\max_{0\leq t\leq T}\frac{1}{L^d}\int v(x,t)d^dx$.  

Several acoustic emission (AE) experiments on small-scale plastic deformations of single crystals, e.g. ice, Cd, Zn, and Cu, report  robust power law scaling in the statistics of the maximum amplitude $A_m$ of acoustic waves emitted during plastic slip avalanches. The probability distribution density of the maximum AE amplitude follows a power-law tail $P(A_m) \sim A_m^{-\mu}$, with an exponent $\mu \approx 2$~\cite{Weiss97,Weiss2000,miguel2001idf,Weiss05,Weiss07,Fressengeas09}. 
Since many slip avalanches are required to obtain good statistics for $P(A_m),$ the deviations in the values of $\mu$ could depend on the experimental resolution. Nevertheless, the exponent is remarkably robust to variations such as loading mode, type of crystal, temperature, forest hardening effect, or plastic anisotropy~\cite{Weiss05,Weiss07,Fressengeas09}. Under certain conditions, it is argued that the maximum amplitude $A_m$ is a measure of the area swept by the fast-moving dislocations~\cite{Weiss97,Weiss2000}. The same power law exponent for the maximum velocity distribution has been observed using high-resolution extensometry~\cite{Weiss07}. 

The picture that emerges from the AE experiments is that crystal deformations due to the intermittent motion of dislocations are characterized by seismic events analogous to those in the dynamics of plate tectonics. Crystal plasticity as a critical phenomenon has also been confirmed by other experimental techniques such as high-resolution extensometry~\cite{Weiss07} and compression of nano-pillars and micro-pillars ~\cite{Dimiduk06,Friedman12}, as well as by numerical simulations of discrete dislocation dynamics~\cite{miguel2001idf,Moretti04,Miguel08,Chan10}. Moreover, the statistics of slip avalanches--i.e. the avalanche size and duration distributions as well as the power spectrum--seem to agree remarkably well with the mean field theory of interface depinning~\cite{Csikor07,TsekenisThesis,TsekenisPreprint,Chan10}. 

\begin{table*}[!t]
\centering
\begin{tabular}{l l l }
\hline\hline
Quantitiy&Form&MFT values\\\hline
Avalanche size distribution&$P(S)\sim S^{-\tau}\mathcal{F}(k^{1/\sigma }S)$&$\tau=\frac{3}{2}$, \hspace{2mm}$\sigma =\frac{1}{2}$\\
Duration distribution&$P(T)\sim T^{-\alpha}\mathcal{G}(k^{\nu z}T)$&$\alpha=2$, \hspace{2mm}$\nu z=1$\\
Maximum velocity distribution &$P(v_m)\sim v_m^{-\mu}\mathcal{H}(k^{\rho}v_m)$&$\mu=2$, \hspace{2mm}$\rho=1$\\
Maximum energy $(E_m\equiv v_m^2)$ distribution &$P(E_m)\sim E_m^{-\lb\mu/2+1/2\rb}\mathcal{H}(k^{\rho}E_m^{1/2})$&$\frac{\mu}{2}+\frac{1}{2}=\frac{3}{2}$, \hspace{2mm}$2\rho=2$\\
Stress-integrated size distribution&$P_{\mathrm{int}}(S)\sim S^{-(\tau+\sigma)}$&$\tau+\sigma=2$\\
Stress-integrated duration distribution&$P_{\mathrm{int}}(T)\sim T^{-(\alpha+\frac{1}{\nu z})}$&$\alpha+\frac{1}{\nu z}=3$\\
Stress-integrated maximum velocity distribution &$P_{\mathrm{int}}(v_m)\sim v_m^{-(\mu+\frac{1}{\rho})}$&$\mu+\frac{1}{\rho}=3$\\
Stress-integrated max energy $(E_m\equiv v_m^2)$ dist. &$P_{\mathrm{int}}(E_m)\sim E_m^{-\lb\frac{1}{2}(\mu+\frac{1}{\rho})+1/2\rb}$&$\frac{1}{2}\lb\mu+\frac{1}{\rho}\rb+\frac{1}{2}=2$\\
Average size vs. duration &$\langle S|T\rangle \sim T^{\frac{1}{\sigma \nu  z}}$&$\sigma \nu  z=\frac{1}{2}$\\
Average max velocity vs. duration &$\langle v_m|T\rangle \sim T^{\frac{\rho}{\nu z}}$&$\frac{\rho}{\nu z}=\frac{\alpha-1}{\mu-1}=1$\\
Average max energy $(E_m\equiv v_m^2)$ vs. duration &$\langle E_m|T\rangle \sim T^{\frac{2\rho}{\nu z}}$&$2\frac{\rho}{\nu z}=2\frac{\alpha-1}{\mu-1}=2$\\
Average max velocity vs. size &$\langle v_m|S\rangle \sim S^{\sigma\rho}$&$\sigma\rho=\frac{\tau-1}{\mu-1}=\frac{1}{2}$\\
Average max energy $(E_m\equiv v_m^2)$ vs. size &$\langle E_m|S\rangle \sim S^{2\sigma\rho}$&$2\sigma\rho=2\frac{\tau-1}{\mu-1}=1$\\\hline\hline
\end{tabular}
\caption{Summary table of the mean-field theory (MFT) exponents and scaling relationships for avalanches statistics in slowly-driven (the driving rate $\tilde c \rightarrow 0$) interfaces near depinning. The first four lines give the power law and cutoff exponents for for the size $S$, duration $T$, maximum velocity $v_m$ and maximum energy $E_m\equiv v_m^2$ observables. The script letters denote universal scaling functions. The first three distributions are shown exactly in Eqs.~(\ref{eq:sizeDist}),~(\ref{eq:durDist2}), and ~(\ref{eq:P(v_m)}) and the fourth follows from a simple change of variables. The parameter $k$ gives the distance to criticality for the system; in Barkhausen noise experiments, it represents the demagnetizing field while in steady-state plasticity scenarios (including earthquakes) it is proportional to the stiffness of the coupling between the system and the driving. In the stress-controlled situation, $k$ can be replaced by $F_c-F,$ where $F$ is the external stress and $F_c$ is the critical stress, and the exponent predictions will remain the same. The second four lines give stress-integrated exponents and the exponents are the expected outcome of plasticity experiments where the external force or stress is increased gradually until failure and avalanches occur along the way. The last five lines give how the size, maximum velocity and maximum energy of an avalanche scale with its duration, and then how the maximum velocity and energy scale with the size. The first is well-known ~\cite{Zapperi98,Durin2000,Colaiori08,Kuntz00} and the last four can be obtained by taking the averages of Eqs.~(\ref{eq:scalingForm}) and~(\ref{eq:sizeScalingForm}).}
\label{table:tbl1}
\end{table*}

Recently, in Ref.~\cite{LeBlanc12} we derived the value of the exponent $\mu$ from a mean field theory of the depinning transition. In this paper, we present the calculations that led to this result in more detail and extend the results to include the effect of a nonzero driving rate and cutoff scaling functions. Although motivated by experiments on crystal plasticity, our results are expected to be broadly applicable to other systems that fall in the mean field interface depinning universality class, such as earthquakes and Barkhausen noise in soft magnets. In Table~\ref{table:tbl1}, we give a summary of the  predictions of the mean field theory, including our new results for the maximum velocity. The predictions we quote are catered to both Barkhausen noise and plasticity experiments (but see the caveats at the end of Section~\ref{sect:OverallDists}), the latter in both steady-state and stress-controlled situations (see the caption to Table~\ref{table:tbl1}).

Some of the distributions that we derive in this paper are relevant to the field of extreme value statistics (EVS), which is of importance to many statistical systems, from natural disasters, e.g. floods~\cite{Katz02}, earthquakes~\cite{Gan83}, snow avalanches~\cite{Mcclung91}, to financial crises~\cite{Poon04} and social reactions to extreme events~\cite{Fortunato06}. Most of these natural systems have a complex evolution characterized by catastrophic, extreme events that appear at intermittent intervals and have a large impact on the long-term behavior of the systems~\cite{Sornette02}. In physics,  determining the likelihood of these rare events is  important to a variety of problems. For example, the low-temperature thermodynamics of spin glasses is governed by the statistics of low-energy states~\cite{Bouchaud97}; the velocity of traveling fronts is selected by the extreme value for a wide class of initial conditions similar to  random bisections that occur during fragmentation processes~\cite{Krapivsky00};  the level density of a noninteracting Bose gas has an asymptotic behavior governed by the three limit laws of classical EVS~\cite{Comtet07} due to a deeper connection between the sum of correlated random variables and extreme values~\cite{Clusel08}; much recent progress on EVS in systems with strong correlation has been achieved in the context of Gaussian height fluctuations of kinetically growing interfaces~\cite{Raychaudhuri01,Majumdar04,Majumdar05,Racz07,Rambeau11}; extreme statistics applies to the largest/smallest eigenvalue of random matrices, also corresponding to the extreme positions occupied in a Coulomb gas with logarithmic interactions~\cite{Dean08}.      
 
Classical extreme value statistics applies to independent, identically distributed random variables $\{x_i\}_{i=1}^N$ with the parent distribution $p(x)$. Under these assumptions, the distribution that the extremum (here taken as the maximum) is less than a given value $m$, namely $C(m)= \textrm{Prob}(\max_{1\le i\le N} x_i<m)$, approaches in the limit of $N\rightarrow \infty$ one of the three types of asymptotic laws depending on the tail of $p(x)$: i) when $p(x)$ decays faster than any power law and is unbounded, e.g. a Gaussian or an exponential function, the EVS is governed by the Fisher-Tippett distribution  $C(m)\sim \exp(\exp(-m))$; ii) when $p(x)$ decays as a power law $p(x)\sim x^{-\alpha}$, with $\alpha>0$, the extreme value distribution approaches the Fr\'{e}chet limit function,  $C(m)\sim \exp(-m^{-\alpha})$; iii) the Weibull distribution function $C(m)\sim \exp(-(-m)^\alpha)$, with $\alpha>0$, determines the EVS class for the random variables that are distributed on a bounded interval. Although a general theory of extreme value statistics for arbitrary correlations and parent distribution $p(x)$ has yet to emerge, substantial progress has been made in several cases where the EVS is exactly solvable for correlated variables. For instance, Berman's theorem says that the EVS of a weakly-correlated Gaussian process is also governed by the Fisher-Tippett distribution~\cite{Berman64,Pickands69}. This corresponds to a power spectrum density that decays with the frequency as $S_f\sim f^{-a}$, with an exponent $0\le a \le 1$. Indeed, the EVS of time records with long-term persistence of Gaussian distributed fluctuations with $a<1$ converges to the Fisher-Tippett distribution~\cite{Eichner06}. The same asymptotic law also determines the distribution of maximum heights of periodic, Gaussian $1/f$-noise~\cite{Racz01,Racz07}, where the maximum is measured relative to the mean. 

On the other hand, the extreme value statistics of a correlated Gaussian process with $a>1$ typically has a simple scaling form with the duration, with a scaling function strongly dependent on several parameters, such as the boundary conditions, the value from which the extremum is measured, as well as other ordering constraints on the time evolution~\cite{Majumdar05,Racz07}. For example, different scaling functions are obtained for the maximum heights of periodic Gaussian interfaces: if the maximum is measured relative to the spatially averaged height, the corresponding EVS is determined by the so-called Airy distribution function~\cite{Majumdar04,Majumdar05,Racz07,Rambeau11}, whereas measuring the maximum relative to the boundary value leads to the Rayleigh distribution~\cite{Rambeau09,Burkhardt07}. We will discuss the extreme value distribution for an avalanche in a mean field theory of interface depinning  known as the Alessandro-Beatrice-Bertotti-Montorsi (ABBM) model~\cite{ABBM90} and will show that this avalanche signal can be viewed as a sequence of strongly-correlated, non-identically distributed, non-Gaussian variables. The probability distribution function (PDF) of the maximum velocity inside avalanches of fixed duration $T$ follows a universal scaling form  $P(v_m|T) = (2v_m T)^{-1/2}F(\sqrt{2v_m/T})$, with a scaling function $F(x)$ that can be derived exactly by a mapping to an equivalent problem of random excursions of Brownian motion in a logarithmic potential.

This paper is structured as follows: In Section II, we review the ABBM model of interface depinning and derive it from a discrete mean field theory. Then in Section III, we outline our methods and use them to demonstrate a derivation of the velocity distribution at a given time in an avalanche of a given duration. We also comment on the relationship between the Ito and Stratonavich interpretations of the square-root multiplicative noise in the model, in particular why they both predict the same parabolic shape for the average velocity as a function of time inside an avalanche. Following that, in Sections IV and V, we use methods similar to those of Section III to derive the distributions of peak velocities for avalanches of given durations and sizes. The main results for fixed durations at the critical point have been recently reported in Ref.~\cite{LeBlanc12}. Here, we present the details of the calculations and extend the work to also include the effects of nonzero sweep-rate $c$ and the maximum velocity statistics for fixed avalanche size. Finally, in Section VI, we derive the peak velocity distribution integrated over all avalanches, as well exact scaling functions describing the tails of size and duration distributions. Concluding remarks and a summary are provided in Section VII.

\section{ABBM model}
Mean-field avalanches in the motion of interfaces near depinning have been extensively studied in the realm of the ABBM model, which was proposed by Alessandro et al. in Refs.~\cite{ABBM90}. The model was initially derived to mimic the dynamics of a single domain wall in soft magnets exhibiting Barkhausen noise, but has become a ~\lq standard model\rq~for interface depinning in the presence of long-ranged forces~\cite{Zapperi98,Colaiori08,Doussal09,Papanikolaou11}. The dynamics of a single interface propagating across a disordered material results from a competition between internal elasticity of the interface, interactions with the quenched disorder experienced by the moving front and an external driving field that is increased at a constant rate.  In the mean-field approximation, the motion of the center-of-mass position of the interface $u(t)=\frac{1}{L^d}\int u(x,t)d^dx$ is given as 
\be\label{eq:ABBM_ut}
\frac{du}{dt}=k\lb \frac{c}{k}t-u\rb+F(u),
\ee
where $c/k$ is the constant pulling rate, $k$ is an elastic coupling to the driving force, and the effective pinning force $F(u)$ is assumed to be Brownian correlated in $u$, i.e. $\langle F(u)\rangle=0$ and
\be
\langle |F(u)-F(u')|^2\rangle = 2D|u-u'|.
\ee
It is important to note that this does not imply long-ranged correlations of disorder in the medium. In fact, under a physically sensible short-ranged disorder correlator $\langle F(x,u)F(x',u')\rangle\sim\delta(x-x')\Delta(u-u'),$ where $\Delta(u)$ is peaked at the origin, the \emph{effective} correlations in the total pinning force $F(\{u(x)\})=\int F(x,u(x))d^dx$ in the center of mass coordinate $u$ are long-ranged because different pieces of the interface move forward at different times and the smaller the fraction that moves, the less the total pinning force changes~\cite{Zapperi98}.
 
The evolution of the average velocity of the interface, defined as $v=du/dt$, can be obtained by differentiating Eq.~(\ref{eq:ABBM_ut}) with respect to $u$. Thus,   
\be\label{eq:ABBM_vu}
\frac{d\tilde v}{du}=-k+\frac{c}{\tilde v}+w(u),
\ee
where $\tilde v(u)\equiv v(t)$ and $w(u) = dF/du$, so that 
\be
\langle w(u)w(u')\rangle = 2D\delta(u-u').
\ee
Eq.~(\ref{eq:ABBM_vu}) can be interpreted as the motion $\tilde v(u)$ of a Brownian particle in a logarithmic potential $\tilde U(u)=ku-c\log(u)$. We will assume that the motion obeys $v(x,t)\geq0$ at all times, so that $u(t)$ is monotonic and we can choose to study the system's evolution in either $u$ or $t.$ By multiplying with $du/dt$ on both sides in Eq.~(\ref{eq:ABBM_vu}), the evolution in time can then be written as 
\be\label{eq:ABBM_vt}
\frac{dv}{dt}=-kv+c+\sqrt{v}\eta(t),
\ee
where the multiplicative noise $\eta(t)\equiv \sqrt{\tilde v}w$ is Gaussian, has zero mean, obeys
\be\label{eq:noiseCorr}
\langle\eta(t)\eta(t')\rangle=2D\delta(t-t'),
\ee
and is interpreted in the Ito sense. The interpretation can be verified by demanding that Eq.~(\ref{eq:ABBM_vt}) predict the correct steady state distribution~\cite{ABBM90}
\be\label{eq:steadyState}
P(v,t\rightarrow\infty)\propto v^{-1+c/D}\exp(-kv/D),
\ee
which can also be derived from Eq.~(\ref{eq:ABBM_vu})~\cite{Zapperi98}. 

Incidentally, we notice that Eq.~(\ref{eq:ABBM_vt}) is identical to the equation satisfied by interest rates in the Cox-Ingersoll-Ross bond pricing model~\cite{CJS85}. A similar $\sqrt{v}$-multiplicative noise process also appears in reaction-diffusion systems driven by internal noise~\cite{Moro04}. The dynamics of this process is characterized by power-law statistics of avalanche sizes and durations, and long-range temporal correlations with power spectrum $S(f)\sim f^{-a},$ where $f$ is the frequency and $a>1$. 

An avalanche of duration $T$ corresponds to an excursion of $v(t)$, which is a path that starts and ends with $v(0)=v(T)=0$ with $v(t)>0$  for $0<t<T$.  We study the statistics of avalanche durations and velocities conditioned on durations from Eq.~(\ref{eq:ABBM_vt}). Alternatively, the avalanche size distributions as well as the velocity statistics conditioned on avalanche sizes can be studied from Eq.~(\ref{eq:ABBM_vu}). In this parametrization, an avalanche of size $S=\int_0^T v(t)dt$ corresponds to an excursion of $\tilde v(u)$ along the coordinate $u$ (integrated velocity). This corresponds to a path that starts and ends at $\tilde v(0)=\tilde v(S)=0$ with $\tilde v(u)>0$  for $0<u<S$. 

\subsection{Other formulations of the mean field theory}

This mean field theory has two other equivalent constructions that provide useful physical pictures. The first uses integer time $t=j \delta t$ and velocity variables and is a Markov process where the number of cells $n_{t+1}$ that fall at time step $t+1$ is drawn from a Poisson distribution
\be\label{eq:PoissonUpdate}
P(n_{j+1}|n_j)=\frac{\rho(n_j)^{n_{j+1}}}{n_{j+1}!}e^{-\rho(n_j)}
\ee
where $\rho(n_t) =\langle n_{t+1}\rangle$. This rule can be derived from the following picture, (see e.g.~Refs.~\cite{Fisher98,Mehta06}): We first divide the interface into $N$ cells of equal volume that are coupled to one another with a mean-field coupling (infinite range) and also each coupled to a driving force that pulls the interface at an average velocity $v_a$.  This means that the total local stress on the $i$'th cell is given by
\be
\tau_i=(J/N)\sum_{j\ne i}^N(u_j-u_i)+K(v_at-u_i)
\ee
where $J$ is the elastic coupling between the cells, $K$ is the  coupling to the driving point and the $u_i$ are the positions of the cells along the direction of motion.

The dynamics is that a cell slips forward by an amount $\delta u_i$ as soon as its local stress is above a threshold $\tau_{s,i}$. This raises the stress on every other cell by an amount $\delta\tau_j=(J/N)\delta u_i$ and decreases cell $i$'s stress by an amount $|\delta\tau_i|=(J+K)\delta u_i.$ Assuming that the stress drops, $\delta \tau_i$, and failure stresses, $\tau_{s,i}$, are distributed very narrowly around the values $\delta\tau$ and $\tau_s$, respectively, each cell's stress will be approximately confined to the interval $\left[\tau_s-\delta\tau,\tau_s\right].$ Then, each cell that did not slip at a given time step will experience a stress increase of
\be
\Delta\tau_t=n_t\frac{J\delta\tau}{N(J+K)}+Kv_a\delta t
\ee
where $n_t$ is the number of cells that slipped at time $t$ and $\delta t$ is the duration of a time step. If we further assume that the stresses of the cells that did not slip at time $t$ are uniformly distributed within $\left[\tau_{s,i}-\delta\tau,\tau_{s,i}\right],$ the cells that fall at time $t+\delta t$ will be the ones with stresses greater than $\tau_{s,i}-\Delta\tau_t.$ This means that each cell has a $\Delta\tau_t/\delta\tau$ chance of slipping at the next time step $t+\delta t$ and that the average number of cells that slip is
\be
\langle n_{t+\delta t}\rangle = N\frac{\Delta \tau_t}{\delta\tau}=\frac{n_tJ}{(J+K)}+\frac{NKv_a\delta t}{\delta\tau}.
\ee
Writing the average number of cells that slip per time step in terms of the pulling velocity as $n_a=v_aN\delta t/\delta u$ and rearranging gives
\be
\rho(n_{t})=\langle n_{t+\delta t}\rangle=\frac{Jn_t+Kn_a}{J+K}
\ee
which has $\langle n_{t+\delta t}\rangle=\langle n_t\rangle=n_a$ as a steady state average.
Since we expect an average of $\rho(n_t)$ cells to slip in the next time period, $n_{t+\delta t}$ is Poisson distributed as in Eq.~(\ref{eq:PoissonUpdate}). 

As also pointed out in Ref.~\cite{Papanikolaou11}, an update rule like Eq.(\ref{eq:PoissonUpdate}) derived in the shell model of the random field Ising model \cite{Sethna93} is equivalent to the ABBM model in the continuum limit. This follows from the Gaussian approximation to the Poisson update rule, which is valid for large $\rho(n_t)$ and is given by
\be
n_{t+\delta t}=\rho(n_t)+\sqrt{\rho(n_t)}\eta_t,
\ee
where $\eta_t$ is a univariate Gaussian random variable with zero mean. Assuming that $K/J$ is small and that terms multiplied by it can be neglected in the fluctuation term when $\rho(n_t)$ is large, we have
\be
n_{t+\delta t}-n_t\approx-\frac{K}{J}(n_t-n_a)+\sqrt{n_t}\eta_t.
\ee
The number of cells that slip during a time step is related to the instantaneous center of mass velocity by $v_t=n_t\delta u/(N\delta t),$ so, multiplying through by the conversion factor and taking the continuum limit in time $\delta t\rightarrow 0$, we arrive at the ABBM equation
\be
\dot v(t) = k\lb\frac{c}{k}-v(t)\rb+\sqrt{2Dv(t)}\xi(t).
\ee
where $k\equiv \frac{K}{J\delta t}$, $c\equiv kv_a,$ $\eta(t)=\eta_t/\sqrt{\delta t},$ and  $2D\equiv \frac{\delta u}{N\delta t^2}.$ The form of the discrete time equation shows that the Ito interpretation is correct. Therefore, this model is equivalent to the ABBM model at large velocities. Notice also that for $K=0,$ the continuum limit of the discrete model's steady state equation
\be
P(n)=\sum_{n'}P(n|n')P(n')\approx\int dn'\frac{(n')^ne^{-n'}}{n!}P(n')
\ee
is satisfied by the ansatz $P(n)=n^{-1},$ which agrees with the $c,k\rightarrow0$ limit of Eq.~(\ref{eq:steadyState}). 

The ABBM model is also the continuum limit of a point process in which the dynamics proceeds in discrete jumps with random waiting times $\tau_j=t_j-t_{j-1}$ in between that are power-law distributed (see e.g. Ref.~\cite{Kaulakys09} and references therein). In particular, the discrete process that leads to Eq.~\ref{eq:ABBM_vt} at the critical point has $P(\tau_j)\sim \tau_j^{-2}.$

For a stationary Poisson process with a constant mean $\lambda$, the waiting time distribution is $P(\tau)=\lambda e^{-\lambda \tau}$, whereas for a modulated Poisson process with $\lambda$ a stochastic variable with a probability distribution $P(\lambda)$, it is
\be\label{eq:poisson}
P(\tau)= \int_0^\infty \lambda e^{-\lambda \tau} P(\lambda)d\lambda.
\ee
The steady state probability of the jumps is $P(n)\sim n^{-1}$,  so $P(\lambda)\sim \lambda^{-1}$, which in turn implies that $P(\tau)\sim \tau^{-1}$ in the continuum. This is consistent with the rule $P(\tau_j)\sim\tau_j^{-2}$ in the discrete case since the probability of a random time being in a waiting interval of duration $\tau$ is proportional to $\tau.$  Since Eq.~(\ref{eq:ABBM_vt}) can be derived as the continuum limit of a sequence of discrete pulses, the Ito interpretation is again seen to be appropriate. 
 
\subsection{Numerics}
In numerical simulations, we use an Euler - Maruyama method to integrate   Eq.~(\ref{eq:ABBM_vt}) with $D=1/2$. This scheme, however, does not preserve the positivity constraint of the solution. Nonetheless, since we are interested in avalanches, i.e. the evolution in between zero crossings, we discard the times when the velocities become negative. An avalanche is started from $v =1$ and evolved until $v\leq 0$ at which time it is declared over and recorded. Then a new avalanche is restarted from $v=1$ with a different realization of the noise. Since the noise is uncorrelated, this closely resembles the reflecting boundary condition at $v=0$ in the steady state evolution of the model. This method is simple and fast, but the disadvantage is that the crude integration rule and boundary conditions introduce inaccuracies for small avalanches ($T\ll 10^3$, $S\ll 10^6$).  Another method tried was directly implementing Eq.~(\ref{eq:PoissonUpdate}) which obtained results consistent with the first method for $k=c=0$, with a somewhat faster convergence to the continuum limit seen in the scaling functions.

\section{Velocity statistics for avalanches of fixed duration}

With the change of variable $x= 2\sqrt{v}$ and Ito's lemma, Eq.~(\ref{eq:ABBM_vt}) can be transformed into an additive noise equation
\be\label{eq:ABBM_Ito}
\frac{dx}{dt}=-\frac{k}{2}x+\frac{2c-D}{x}+\xi(t).
\ee
The problem is thus mapped to an over-damped Brownian motion of a particle confined to the right half plane in a potential $V(x)=\frac{1}{2}kx^2-(2c-D)\log(x)$. Near criticality, $k\rightarrow0$ and the problem reduces to Brownian motion in a logarithmic trap. For an initial condition $x(0)=0$, the particle may execute an excursion with $x(t)>0$  for a duration $T$ until its position returns to the origin for the first time. This excursion corresponds to an avalanche in the $v(t)$ during the interface propagation. In the following, we  calculate the distribution of the instantaneous displacement during an excursion of size $T$, namely $P(x,t|T)$, and, by a corresponding transformation of variables, the conditional distribution of velocities in avalanches of fixed durations, $P(v,t|T)$. This allows us to determine the avalanche shape.   

Following a standard path-integral method, the probability distribution for the process defined by Eq.~(\ref{eq:ABBM_Ito}) can be written as a path integral
\bea
P(x,t)&\propto&\int \mathcal{D}y(\tau)\delta(x-y(t))\nonumber\\&\times&\left\langle\delta\lb\frac{dy}{dt}+\frac{dV(y(t)}{dy}-\xi(t)\rb \right\rangle_\xi
\eea
where $V(x)$ is as above and the noise average is performed with respect to the Gaussian distribution $P_\xi\propto \exp\lb-\frac{1}{4D}\int \xi(t)^2dt\rb$ that produces, e.g. Eq.~(\ref{eq:noiseCorr}). Following Ref.~\cite{Majumdar05}, we demand that the path is an excursion of duration $T$ by fixing the endpoints $x(0)=x(T)=\eps,$ where $\eps$ is some small value, later taken to zero. Also, by applying the positivity constraint that $x(t)>0$ for $0<t<T$, we have that 
\bea\label{eq:path_int_Pxt}
P(x,t|T)&=&\lim_{\eps\rightarrow0}\frac{1}{Z_\eps(T)}\int_{y(0)=\eps}^{y(T)=\eps}\mathcal{D}y(\tau)e^{-\int_0^Td\tau L_E(y,\dot y)}\nonumber\\
&\times& \delta(y(t)-x)\prod_{0\le \tau\le T}\Theta(y(\tau)).
\eea
The noise average has been performed, resulting in the Lagrangian $L_E\left(x,\frac{dx}{dt}\right)=\frac{1}{4D}\left(\frac{dx}{dt}+\frac{dV(x)}{dx}\right)^2$ and the product of theta functions enforces the positivity constraint. The normalization factor integrates over all excursions regardless if they go through $x$ at time $t$ and is therefore given by
\bea
Z_\eps(T) &=& \int_{x(0)=\eps}^{x(T)=\eps}\mathcal{D}x(\tau)e^{-\int_0^Td\tau L_E(x,\dot x)}\nonumber\\
&\times&\prod_{0\le \tau\le T}\Theta(x(\tau)).
\eea 
The equivalent real-time Lagrangian, obtained by transforming the imaginary time coordinate $t$ to $\tau=-it$ and factoring out a $-1$ is $L(x,\dot x)=-\frac{1}{4D}\left(-i\dot{x}+V'(x)\right)^2$, where $\dot x=\frac{dx}{d\tau}$ and $V'(x)=\frac{dV}{dx}$. The canonical momentum corresponding to $L(x,\dot x)$ is then $p=\frac{\partial L}{\partial \dot{x}}=\frac{i}{2D}(-i\dot{x}+V'(x))$, and the real time Hamiltonian $\hat H=p\dot{x}-L=Dp^2-ipV'(x)$. Therefore, the evolution of the Brownian particle in a potential $V(x)$ can be determined from its quantum analog that satisfies $i\partial_\tau \psi = \hat H\psi$. Replacing $p=-i\partial_x$ and $\tau= -it $, the probability $P(x,t)$ that the Brownian particle is at position $x$ at time $t$ satisfies the Fokker-Planck equation
\be\label{eq:PF_Ito_1}
\frac{\partial P}{\partial t}= \frac{\partial}{\partial x}\left(V'(x)P\right)+D\frac{\partial^2 P}{\partial x^2},
\ee
which must be solved with an absorbing boundary condition at the origin $P(0,t)=0$ in order to enforce the positivity constraint. For $k=0$, Eq.~(\ref{eq:PF_Ito_1}) reduces to 
\be\label{eq:PF_Ito_2}
\frac{\partial P}{\partial t}= \frac{\partial}{\partial x}\left(\frac{D-2c}{x}P\right)+D\frac{\partial^2 P}{\partial x^2},
\ee
which can be solved exactly.

By a generic separation of variables $P(x,t)=e^{-E t}f(x)$, Eq.~(\ref{eq:PF_Ito_2}) is reduced to the eigenvalue problem
\be
\frac{d^2f}{dx^2}+\frac{1-2\tilde c}{x}\frac{df}{dx}+\lb\frac{E}{D}-\frac{1-2\tilde c}{x^2}\rb f=0,
\ee
where $\tilde c=c/D$. This is a modified version of the Bessel equation and  the particular solution that is well-behaved at the origin is
\be
f(x)=x^{\tilde c}J_{1-\tilde c}\left(\sqrt{\frac{E}{D}}x\right)
\ee
where $J_{1-\tilde c}(x)$ is the Bessel function of the first kind \cite{AandS}. We have assumed $0\leq\tilde c<1$ since this is the range in which the model exhibits avalanche behavior (otherwise the origin is inaccessible in the steady state, as can be seen in Eq.~(\ref{eq:steadyState})).
We can now proceed to calculate the $P(x,t)$. For this, we interpret the above path integrals as a matrix element between two functions defined on $[0,\infty)$ given by
\be
\langle h|e^{-\hat H_1(t-t_0)}|g\rangle_{w}=\int_0^\infty h(x)e^{-\hat H_1(t-t_0)}g(x)dx.
\ee
where the Hamiltonian is $\hat H_1=-D\partial_x^2-\partial_x V_1'(x)$ with the potential $V_1(x)= -(2c-D)\log(x)$ and an absorbing boundary condition at $x=0$. We write the eigenfunctions $f_k(x)=x^{\tilde c}J_{1-\tilde c} (kx)$ with the eigenvalues defined as $E_k=D k^2$.
Using the inverse Hankel transform of $g(x)x^{-\tilde c} = \int_0^\infty k dk J_{1-\tilde c}(kx) \tilde g_{1-\tilde c}(k)$, the above expression becomes equivalent to 
\bea\label{eq:fg}
&&\langle h|e^{-\hat H_1(t-t_0)}|g\rangle_{w} = \int_0^\infty dxh(x)\times\nonumber\\
&\times& \int_0^\infty kdke^{-Dk^2(t-t_0)}x^{\tilde c}J_{1-\tilde c}(kx)\tilde g_{1-\tilde c}(k), 
\eea
where $\tilde g_{1-\tilde c}(k)= \int_0^\infty J_{1-\tilde c}(kx)g(x)x^{1-\tilde c}dx$. 

Hence, the probability of being at position $x$ and time $t$ during an excursion of duration $T$, $P(x,t|T)$, can be defined using the matrix elements as  
\be\label{eq:Pxt_T}
P(x,t|T)= \lim_{\eps\rightarrow0}\frac{\langle\epsilon |e^{-\hat H_1 (T-t)}| x\rangle_w\langle x|e^{-\hat H_1 t}|\epsilon\rangle_w}{\langle\epsilon|e^{-\hat H_1 T}|\epsilon\rangle_w},
\ee
where $\langle\epsilon|e^{-\hat H_1 T}|\epsilon\rangle_{w} = Z_\eps(T)$. Thus, from Eq.~(\ref{eq:fg}) with $h(x)=\delta(x-z)$ and $g(x)=\delta(x-\eps)$, we have 
\bea\label{eq:zt_e0}
\langle z|e^{-\hat H_1 t}|\eps\rangle_w&=&\int_0^\infty dx\delta(x-z)\int_0^\infty kdk x^{\tilde c}J_{1-\tilde c}(kx)e^{-Dk^2t}\nonumber\\
&\times&\int_0^\infty y^{1-\tilde c}dyJ_{1-\tilde c}(ky)\delta(y-\eps)\\\nonumber
&=&\eps^{1-\tilde c}z^{\tilde c}\int_0^\infty k dk e^{-Dk^2t}J_{1-\tilde c}(k\eps)J_{1-\tilde c}(kz),
\eea
and, in the limit of $\eps\ll 1$, the above expression can be expanded to leading order  
\bea
\langle z|e^{-\hat H_1 t}|\eps\rangle_w
&\approx&\frac{\eps^{2(1-\tilde c)}z^{\tilde c}}{2^{1-\tilde c}\Gamma(2-\tilde c)}\int_0^\infty e^{-Dk^2t}J_{1-\tilde c}(k z)k^{2-\tilde c}dk\nonumber\\
&\approx&\left(\frac{\eps}{2}\right)^{2(1-\tilde c)}\frac{z(Dt)^{-2+\tilde c}}{2\Gamma(2-\tilde c)}e^{-\frac{z^2}{4 Dt}}
\eea
where $\Gamma(z)$ is the gamma function.
\begin{figure}[!t]
\includegraphics[trim=1cm 0 0 0,width=0.45\textwidth]{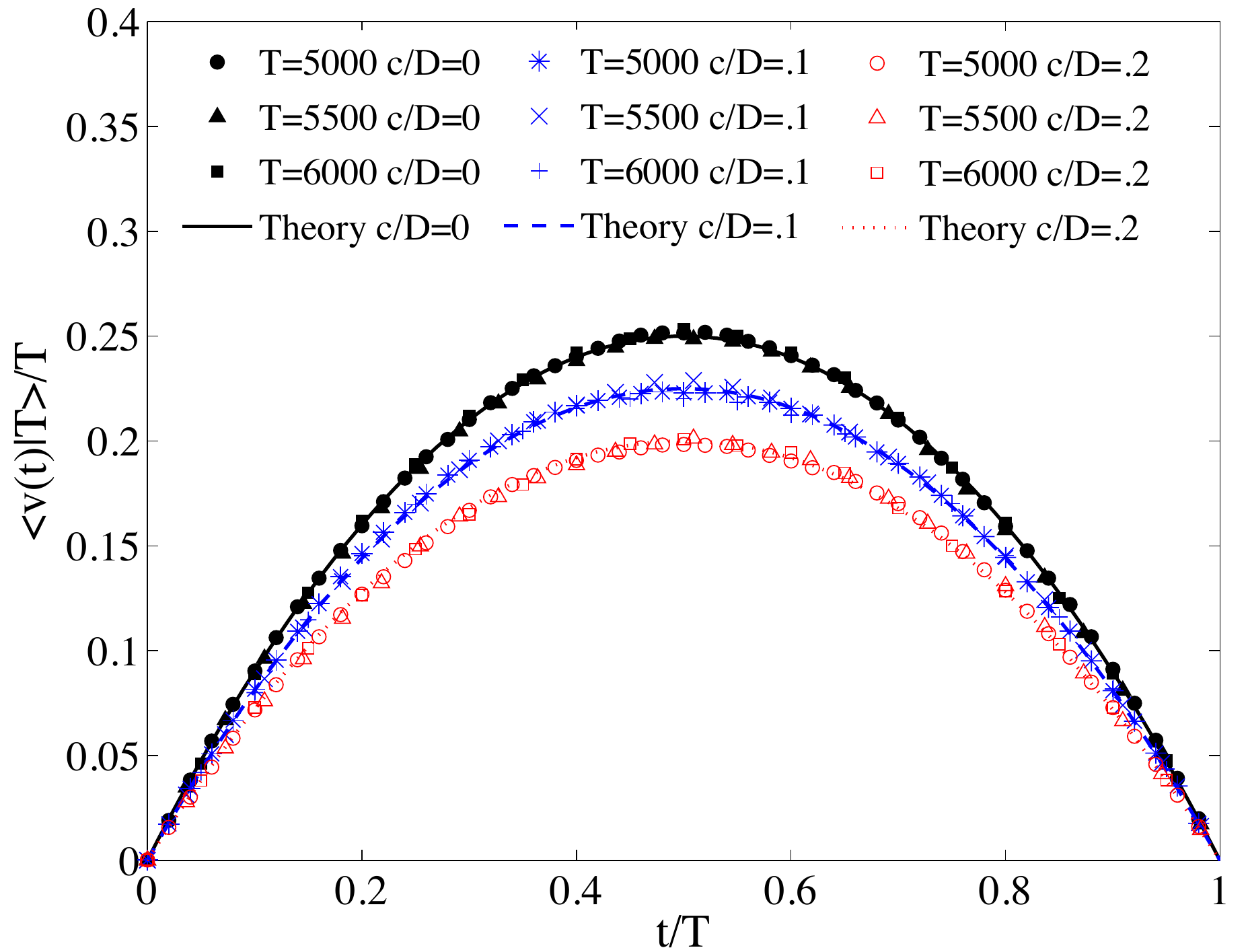}
\caption{ (Color online) Numerical results for the instantaneous velocity $v(t)$ at time $t$ averaged over many avalanches of duration $T$ and rescaled to collapse to the form of Eq.~(\ref{eq:Pvt_T}) with $n=1.$ Results for three different values of the dimensionless driving rate $\tilde c=c/D$ are shown. (In the numerics, the driving rate $c$ is varied while the fluctuation strength is fixed at $D=1/2$.) The parabolic shapes with $\tilde c$-dependent height predicted by Eq.~(\ref{eq:shapes}) for $n=1$ match well with the numerics.} \label{fig:1}
\end{figure}

In a similar way, we obtain the other transition amplitude, to the leading order in $\eps$, as 
\bea
&&\langle \eps|e^{-\hat H_1 (T-t)}|z\rangle_{w}\nonumber\\
&\approx&\frac{\eps z^{1-\tilde c}}{2^{1-\tilde c} \Gamma(2-\tilde c)}\int_0^\infty e^{-D k^2(T-t)}J_{1-\tilde c}(k z)k^{2-\tilde c}dk\nonumber\\
&\approx&\left(\frac{z}{2}\right)^{2(1-\tilde c)}\frac{\eps[D(T-t)]^{-2+\tilde c}}{2\Gamma(2-\tilde c)}e^{-\frac{z^2}{4 D(T-t)}}.
\eea
The normalization constant $Z_\eps(T)$ can also be determined with $h(x)=\delta(x-\eps)$ and $g(x)=\delta(x-\eps)$, and is given by 
\bea\label{eq:ZE_T}
Z_\eps(T)&=&\eps\int_0^\infty\left[J_{1-\tilde c}(k\eps)\right]^2e^{-Dk^2T}kdk\nonumber\\
&=&\frac{\eps}{2D T}I_{1-\tilde c}\lb\frac{\eps^2}{2D T}\rb e^{-\frac{\eps^2}{2DT}}\nonumber\\
&\approx&\frac{(\eps/2)^{3-2\tilde c}}{(DT)^{2-\tilde c}\Gamma(2-\tilde c)},
\eea
where $I_{1-\tilde c}(x)$ is the modified Bessel function of the first kind and $\tilde c<2$ \cite{AandS}. 
Combining these final expressions with Eq.~(\ref{eq:Pxt_T}), we arrive at 
\bea\label{eq:P_RE_fin}
P(x,t|T)&=&\frac{(x/2)^{3-2\tilde c}}{\Gamma(2-\tilde c)}\left(\frac{T}{Dt(T-t)}\right)^{2-\tilde c}\nonumber\\
&\times& \exp\left(-\frac{x^2T}{4Dt(T-t)}\right).
\eea
Finally, the distribution of velocities at a given time $t$ inside an avalanche of duration $T$ follows by a change of variable $v=x^2/4$ and is given as 
\bea\label{eq:Pvt_T}
&P(v,t|T) = \frac{v^{1-\tilde c}}{\Gamma(2-\tilde c)}\left(\frac{T}{Dt(T-t)}\right)^{2-\tilde c} \exp\left(-\frac{vT}{Dt(T-t)}\right).
\eea
From Eq.~(\ref{eq:Pvt_T}), we determine the profile of the $n$-th moments to be 
\be \label{eq:shapes}
\langle v^n(t)|T\rangle = \frac{\Gamma(n+2-\tilde c)}{\Gamma(2-\tilde c)}\lb\frac{Dt(T-t)}{T}\rb^{n},
\ee
with $n=1$ corresponding to the avalanche shape $\langle v(t)|T\rangle$ (see Fig.~\ref{fig:1}). We notice that the avalanche shape and higher moment profiles are independent of the driving rate $\tilde c$ up to a nonuniversal prefactor. We have also verified this numerically as shown in Fig.~(\ref{fig:1}). 

This method of obtaining Eq.~(\ref{eq:Pvt_T}) has the advantage of being more intuitive thanks to the correspondence with excursions and we will find that this method continues to be very useful when we compute the statistics of avalanche maxima below. However, Eq.~(\ref{eq:Pvt_T}) can also be obtained  from the exact solutions to the untransformed $k\neq0$ Fokker-Planck equation~\cite{ABBM90}
\be
\partial_tP=D\partial^2_v(vP)+\partial_v((kv-c)P),
\ee
which were obtained by Feller~\cite{Feller51}. The exact propagator with an absorbing boundary condition at $v=0$ is given by
\bea \label{eq:exactProp}
P(v,t,v_0,0)&=&\lb\frac{\frac{k}{D}\exp\lb-\frac{\frac{k}{D}(v+v_0e^{-kt})}{1-\exp(-kt)}\rb}{1-e^{-kt}}\rb\\\nonumber&\times&\lb\frac{v}{v_0e^{-kt}}\rb^{(\tilde c-1)/2}I_{1-\tilde c}\lb\frac{2 k\sqrt{v_0ve^{kt}}}{D(e^{kt}-1)}\rb,
\eea
assuming that $0<\tilde c<1.$ Repeating the steps above gives
\begin{eqnarray}
P(v,t|T) &=& \frac{v^{1-\tilde c}}{\Gamma(2-\tilde c)}\left(\frac{\frac{k}{D}\left(1-e^{-kT}\right)}{\left(1-e^{-kt}\right)\left(1-e^{-k(T-t)}\right)}\right)^{2-\tilde c} \nonumber\\&\times&\exp\left(-\frac{\frac{k}{D}v\left(1-e^{-kT}\right)}{\left(1-e^{-kt}\right)\left(1-e^{-k(T-t)}\right)}\right)
\end{eqnarray}
and
\bea
\langle v^n(t)|T\rangle &=&\nonumber \frac{\Gamma(n+2-\tilde c)}{\Gamma(2-\tilde c)}\\&\times&\lb \frac{D\lb1-e^{-kt}\rb\lb1-e^{-k(T-t)}\rb}{k\lb1-e^{-kT}\rb}\rb^{n},
\eea
which reduce to Eqs.~(\ref{eq:Pvt_T}) and~(\ref{eq:shapes}) when $k\rightarrow0.$

Interestingly, the same avalanche shape is obtained when  Eq.~(\ref{eq:ABBM_vt}) is interpreted in the Stratonovich sense, as derived in Ref.~\cite{Papanikolaou11}. In this interpretation, the $x=2\sqrt{v}$ transformation of the equation of motion is
\be\label{eq:ABBM_Strat}
\frac{dx}{dt}=-\frac{k}{2}x+\frac{2c}{x}+\xi(t),
\ee
which is a free Brownian motion in the limit $k=c=0.$ Using the same method as above, the conditional distribution of velocities in avalanches of duration $T$ is (assuming $k=0$), 
\be
P_{\mathrm{Strat.}}(v,t|T)=\frac{v^{1/2-\tilde c}}{\Gamma(3/2-\tilde c)}\lb\frac{T}{Dt(T-t)}\rb^{3/2-\tilde c}e^{-\frac{vT}{Dt(T-t)}}
\ee
and the $n$-th moments are
\be
\langle v^n(t)|T\rangle_{\mathrm{Strat.}}= \frac{\Gamma(n+3/2-\tilde c)}{\Gamma(3/2-\tilde c)}\lb\frac{Dt(T-t)}{T}\rb^{n},
\ee
which is indeed identical to the results from the Ito case up to a non-universal, constant pre-factor. In Ref.~\cite{Papanikolaou11}, it is conjectured that the Stratonovich interpretation would give the same results as the Ito interpretation for all universal quantities, but this is not the case. Notice that Eq.~(\ref{eq:ABBM_Strat}) can be transformed into Eq.~(\ref{eq:ABBM_Ito}) by a simple parameter shift $c\rightarrow c-D/2.$ Therefore, any quantity that has $c$-dependence will be different in the two interpretations. This includes the avalanche size scaling exponent $\tau$ and the duration scaling exponent $\alpha$ which have the values $\tau=(3-\tilde c)/2$ and $\alpha = 2-\tilde c$ (see Refs.~\cite{Colaiori08,Zapperi98} and below). The Stratonovich interpretation predictions for these exponents are then $\tau_S = 5/4-\tilde c/2$ and $\alpha_S = 3/2-\tilde c.$ The fact that the Stratonovich interpretation gives the same parabolic avalanche shapes and higher moment profiles is only a reflection of the fact that those quantities do not depend on the driving rate $c$ (except in an overall prefactor).

\section{Maximum velocity statistics for avalanches of fixed duration}
\begin{figure}[!t]
\includegraphics[trim=1cm 0 0 0,width=0.45\textwidth]{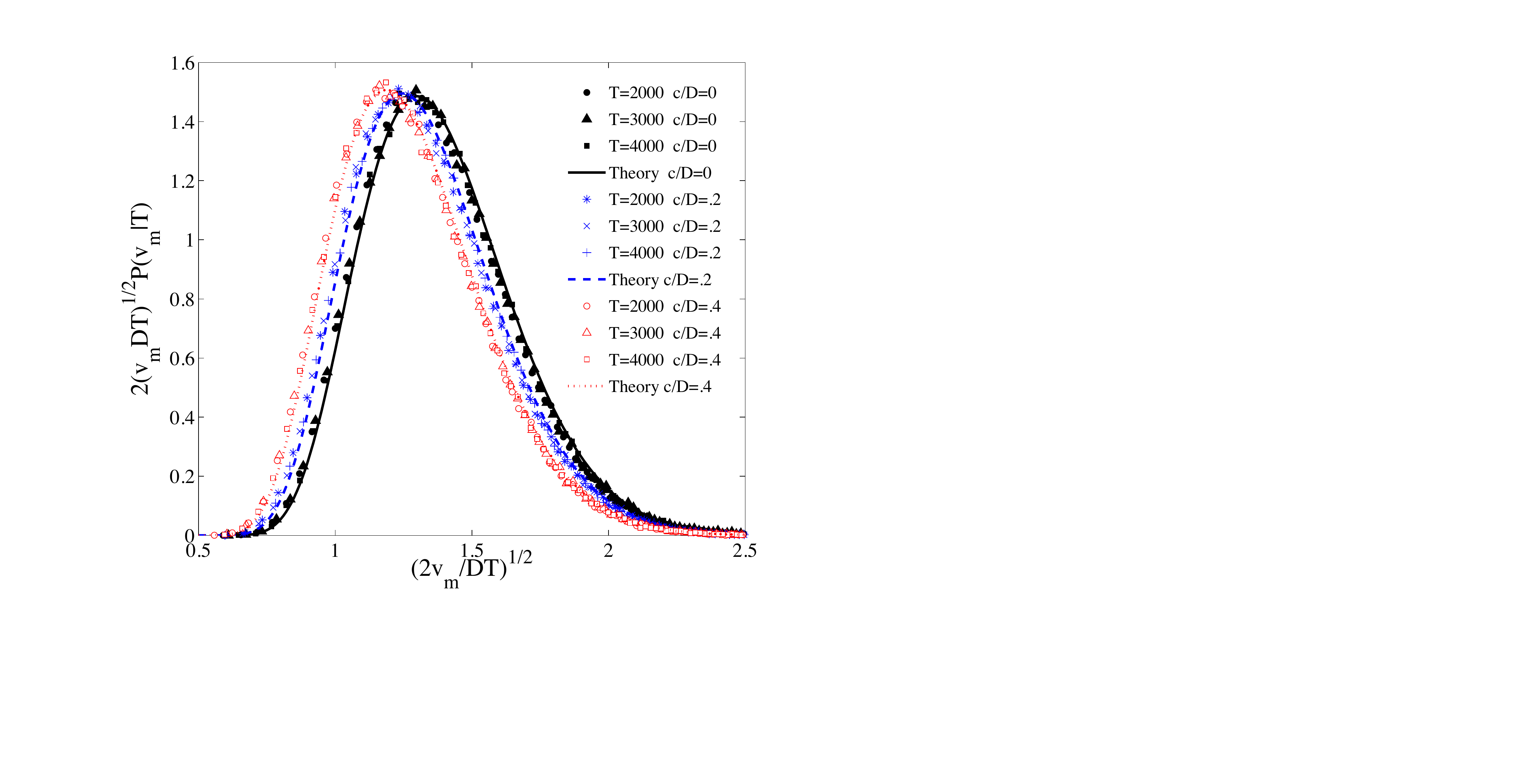}
\caption{ (Color online) Scaling collapses of numerical results of $P(v_m|T),$ the distribution of maximum velocities for avalanches of duration $T$, onto the form given by Eq.~(\ref{eq:scalingForm}) performed for several values of the dimensionless driving rate $\tilde c=c/D.$ (In the numerics, the driving rate $c$ is varied while the fluctuation strength is fixed at $D=1/2$.) The corresponding scaling functions $F(x)$ given by Eq.~(\ref{eq:scalingFunc}) fit very well with the numerics.}\label{fig2}
\end{figure}

We now determine the extreme value distribution of the maximum velocity inside avalanches of fixed duration $T$, for $k=0$ and nonzero $c$. Using the mapping to the random excursions, we first determine the statistics of the maximal displacements in excursions over a fixed interval. Let us consider an  excursion $\{x(t)\}_t$ in the time interval $0\le t\le T$. The probability distribution at a fixed time $t$ is given by $P(x,t|T)$ from Eq.~(\ref{eq:P_RE_fin}), which can be written  
\be
P(x,t|T) = \frac{1}{\sigma_t}Q_{RE}\left(\frac{x}{2\sigma_t}\right),
\ee
with $Q_{RE}(y)=y^{3-2\tilde c}e^{-y^2}/\Gamma(2-\tilde c)$ and $\sigma_t = \sqrt{Dt(T-t)/T}$. Notice that because $\sigma_t$ is different for every point inside the excursion, it follows that the variables $x(t)$ are not identically distributed. Furthermore, the individual distributions are non-Gaussian for small $x$, but converge to a Gaussian right tail for large enough $x$. The maximum displacement in an excursion is defined as 
\be
M = \max_{0\le t\le T} x(t),
\ee
with a probability distribution determined from the properties of $P(x,t|T)$. This follows from the fact that the probability of the maximum being less than a certain value $M<x_m$ is the same as the probability that at every instant during the excursion the displacement $x(t)$ is less than $x_m$. The path integral representing this cumulative probability is 
\bea \label{eq:Cxm_T}
C_\eps(x_m|T)&=&\frac{1}{Z_\eps(T)}\int_{x(0)=\epsilon}^{x(T)=\eps} \mathcal{D}x(t)e^{-\int_0^T L_E(x,\dot x)dt}\nonumber\\
&\times&\prod_{0\le t\le T}\Theta(x(t))\Theta(x_m-x(t)),
\eea
where $L_E(x,\dot x)=\frac{1}{4D}\left(\dot x+\frac{D-2c}{x}\right)^2$ and the product over the Heaviside functions selects the excursion paths that satisfy the constraint that the position at any time is less than $x_m$. The normalization constant $Z_\eps(T)$, corresponding to the path integral over all excursions, was calculated previously in Eq.~(\ref{eq:ZE_T}). At the end of the calculation, we take the limit as $\eps\rightarrow 0$, so that the path starts and ends at the origin, staying between $0$ and $x_m$ during an avalanche of duration $T$. On the other hand, $C_\eps(x_m|T)$ is also the probability that $M<x_m$, namely
\bea 
C_\eps(x_m|T) = \int_{-\infty}^{x_m}dM P_\eps(M|T),
\eea
thus, the PDF $P_\eps(x_m|T)$ follows directly by a differentiation of $C_\eps(x_m|T)$ with respect to $x_m$. 

Using the quantum analogue, the path integral in Eq.~(\ref{eq:Cxm_T}) is equal to a transition amplitude between the position eigenstate $|\eps\rangle$ and itself, and thus can be computed by an expansion in energy eigenfunctions:
\bea\label{eq:cum}
C_\eps(x_m|T)&=&\frac{\langle\eps|e^{-\hat H_1 T}|\eps\rangle_{b}}{\langle\eps|e^{-\hat H_1 T}|\eps\rangle_{w}}, 
\eea
with 
\bea\label{eq:box}
\langle\eps|e^{-\hat H_1 T}|\eps\rangle_{b} = \int_0^{x_m} dx \delta(x-\eps) e^{-\hat H_1 T}\delta(x-\eps),
\eea
where the Hamiltonian $\hat H_1$ has square-well (box) boundary conditions at $x=0$ and $x=x_m$. With these boundary conditions, the Hamiltonian has a discrete spectrum of eigenfunctions given by 
\bea\label{eq:fn_box}
f_n(x) = x^{\tilde c}J_{1-\tilde c}\lb\frac{\lambda_n x}{x_m}\rb,
\eea
with eigenvalues $E_n = D\lambda_n^2/x_m^2$, where $\lambda_n$ is the n'th zero of the Bessel function $J_{1-\tilde c}(x)$. The $\delta$-function in Eq.~(\ref{eq:box}) can be expanded in this basis as 
\bea
\delta(x-\eps) = \frac{2\eps^{1-\tilde c}}{x_m^2}\sum_n \frac{J_{1-\tilde c}\lb\frac{\lambda_n\eps}{x_m}\rb}{\lb J_{2-\tilde c}(\lambda_n)\rb^2}f_n(x),
\eea
with $f_n(x)$'s defined in Eq.~(\ref{eq:fn_box}). Inserting this expansion into the definition of the matrix element from Eq.~(\ref{eq:box}), we obtain    
\bea\label{eq:Cxm_T_2}
&&\langle\eps|e^{-\hat H_1 T}|\eps\rangle_{b}=2\eps\sum_{n=1}^\infinity\left[\frac{J_{1-\tilde c}\lb\frac{\lambda_n\eps}{x_m}\rb}{x_m J_{2-\tilde c}(\lambda_n)}\right]^2e^{-\frac{D \lambda_n^2T}{x_m^2}}.
\eea
Finally, inserting the solution of $Z_\eps(T)$ from Eq.~(\ref{eq:ZE_T}) into Eq.~(\ref{eq:cum}), and taking the limit of $\eps\rightarrow 0$, we arrive at the expression for the extreme value statistics of the random displacements $C(x_m|T)=\lim_{\eps\rightarrow0}C_\eps(x_m|T)$, namely 
\bea\label{eq:Cxm_T_3}
C(x_m|T)&=&\frac{2^{\tilde c}(DT)^{2-\tilde c}}{\Gamma(2-\tilde c)x_m^{4-2\tilde c}}\sum_{n=1}^\infty\frac{\lambda_n^{2(1-\tilde c)}}{[J_2(\lambda_n)]^2}e^{-\frac{\lambda_n^2 DT}{x_m^2}}.
\eea

The distribution $P(x_m|T)$ of the maximum displacements follows from the cumulative distribution as $P(x_m|T)=\partial_{x_m}C(x_m|T)$. The corresponding PDF of the maximal avalanche velocities is obtained by the change of variable $x_m = 2\sqrt{v_m}$ and the transformation 
\be
P(v_m|T)=P(x_m|T)\left|\frac{dv_m}{d x_m}\right|^{-1}.
\ee
It follows that the PDF $P(v_m|T)$ is given by
\be \label{eq:scalingForm}
P(v_m|T)=\frac{1}{\sqrt{2D v_mT}}F\lb \sqrt{\frac{2v_m}{D T}}\rb
\ee
with the scaling function given by $F(x)=$
\be\label{eq:scalingFunc}
\frac{2^{\tilde c}}{\Gamma(2-\tilde c)}\frac{1}{x^{5-2\tilde c}}\sum_{n=1}^\infty\frac{\lambda_n^{2-2\tilde c}}{[J_{2-\tilde c}(\lambda_n)]^2}\lb\frac{\lambda_n^2}{x^2}-(4-2\tilde c)\rb e^{-\frac{\lambda_n^2}{2x^2}}.
\ee

We have numerically verified Eq.~(\ref{eq:scalingForm}) for different values of $\tilde c$ and $k=0$. Our collapsed distributions as shown in Fig.~\ref{fig2} are in excellent agreement with the analytically predicted scaling function from Eq.~(\ref{eq:scalingFunc}).

\section{Maximum velocity statistics for fixed avalanche size}
\begin{figure}[!t]
\includegraphics[trim=1cm 0 0 0,width=0.45\textwidth]{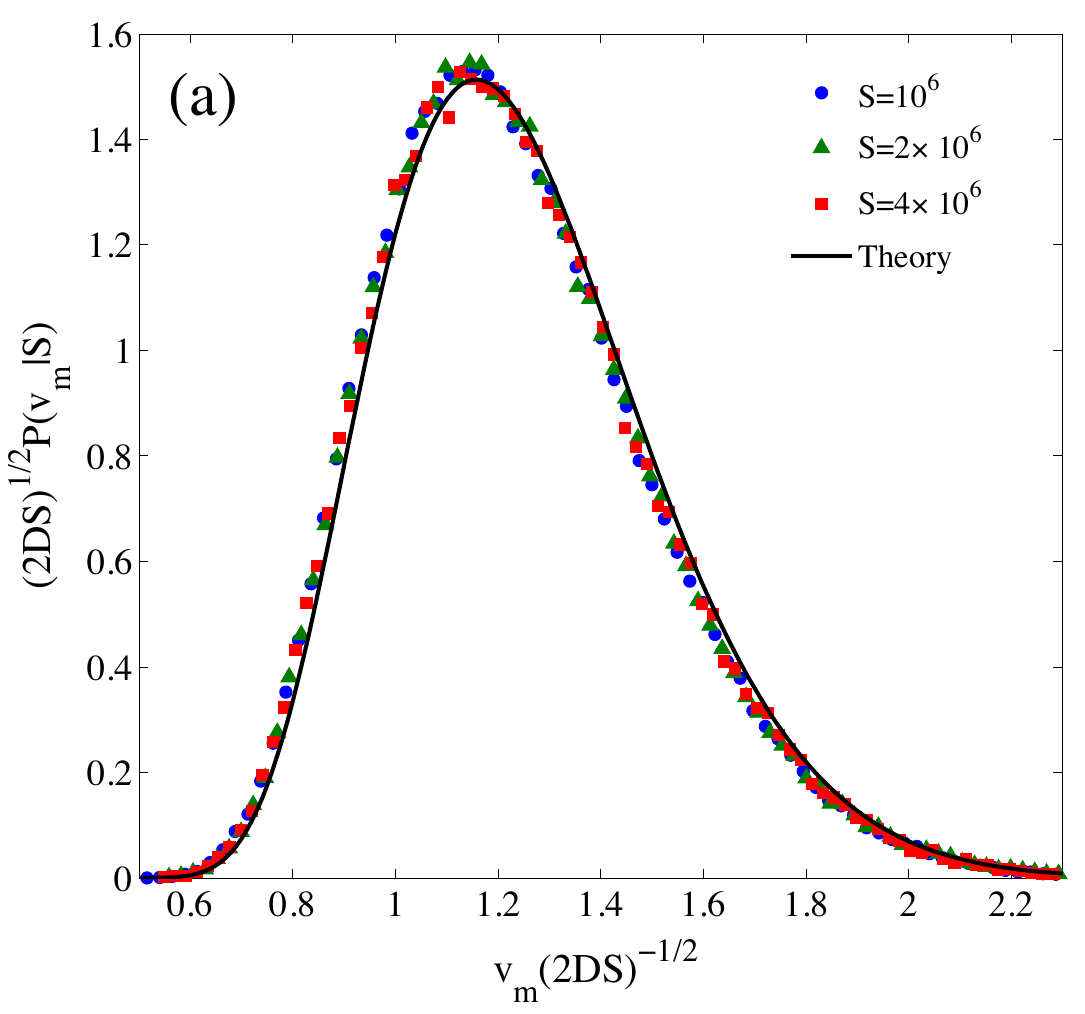}
\includegraphics[trim=1cm 0 0 0,width=0.45\textwidth]{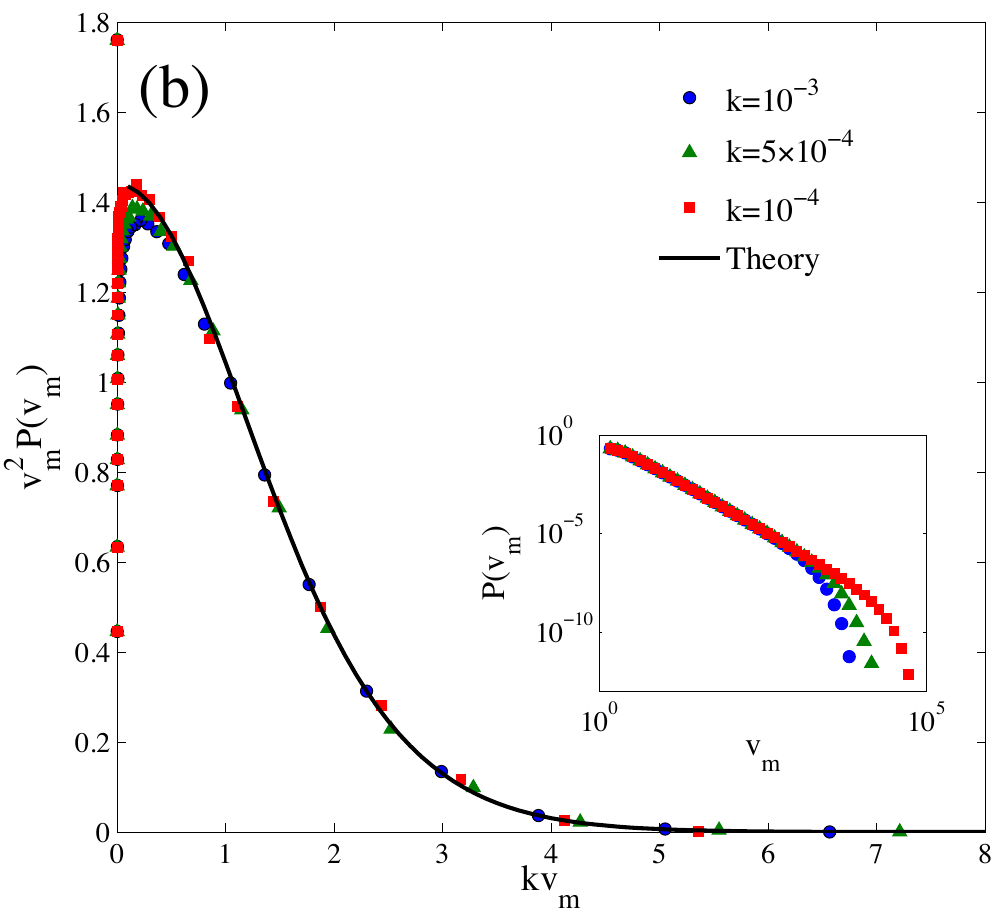}
\caption{ (Color online) (a) Scaling collapse of $P(v_m|S),$ the distribution of maximum velocities for avalanches of  size $S,$ onto the form of Eq.~(\ref{eq:sizeScalingForm}). The rescaled distributions approach the analytical scaling function from Eq.~(\ref{eq:sizeScalingFunc}) for sufficiently large avalanche sizes. (b) Scaling collapse for the overall maximum velocity distribution, $P(v_m)$ for quasistatic driving ($c=0$) and several different values of $k$ (see the caption of Table~\ref{table:tbl1} for the physical definition) to the form of Eq.~(\ref{eq:P(v_m)}). The scaling function plotted is $F(x)=C x^2\sinh^{-2}(x)$ where the constant $C=1.44$ is adjusted to fit near the tail and is proportional to that given in Eq.~(\ref{eq:P(v_m)}) with $D=1/2$. Adjustment of this non-universal constant factor is required since the part of the distribution near the origin that deviates from scaling alters the overall normalization. The inset shows the same maximum velocity distributions before rescaling.} \label{fig:3}
\end{figure}

Using the same technique as in the previous section, we determine the distribution of maximum events in avalanches of a given size. Here, we consider Eq.~(\ref{eq:ABBM_vu}) with $c=0$ and nonzero $k$, which corresponds to a biased Brownian motion 
\be\label{eq:ABBM_vu_RW}
\frac{d\tilde v}{du}=-k + w(u),
\ee
where $\langle w(u)w(u')\rangle = 2D\delta(u-u')$. 

The cumulative distribution $C(v_m|S)$ conditioned on fixed avalanche sizes is defined by the path integral  
\bea \label{eq:Cvm_S}
C_\eps(v_m|S)&=&\frac{1}{Z_\eps(S)}\int_{\tilde v(0)=\epsilon}^{\tilde v(S)=\eps} \mathcal{D}\tilde v(u)e^{-\frac{1}{4D}\int_0^S(\dot{\tilde v}(u)+k)^2du}\nonumber\\
&\times&\prod_{0\le u\le S}\Theta(\tilde v(u))\Theta(v_m-\tilde v(u)),
\eea
where the normalization factor is the path integral over the unconstrained excursions, 
\bea\label{eq:Z_S}
Z_\eps(S)&=&\int_{\tilde v(0)=\eps}^{\tilde v(S)=\eps} \mathcal{D}\tilde v(u)e^{-\frac{1}{4D}\int_0^S(\dot{\tilde v}(u)+k)^2du}\nonumber\\
&\times&\prod_{0\le u\le S}\Theta(\tilde v(u)).
\eea
Instead of duration $T$ as in the previous calculation, the value of $u$ at the end of an avalanche is the slip size of the avalanche $S=\int_0^T v(t)dt$. The path integral in Eq.~(\ref{eq:Z_S}) is equivalent to the transition amplitude between the position eigenstate $|\eps\rangle$ and itself, and thus can be computed by an expansion in energy eigenfunctions:
\bea\label{eq:Z_u}
Z_\eps(S)&=&\langle\eps|e^{-\hat H_2 S}|\eps\rangle_{w}, 
\eea
where $\hat H_2 = -D\partial_{\tilde v}^2-k\partial_{\tilde v}$ with an absorbing boundary condition at $\tilde v=0$. The eigenfunctions for this Hamiltonian are 
\begin{equation}
f_q(\tilde v)=\sqrt{\frac{2}{\pi}}\sin(q\tilde v) e^{-k\tilde v/2D},
\end{equation}
with eigenvalues satisfying $E_q=D\lb\lb\frac{k}{2D}\rb^2+q^2\rb$ for $0<q<\infty$. Therefore, the $Z_\eps(S)$ from Eq.~(\ref{eq:Z_u}) becomes 
\be
Z_\eps(S)=\int_0^\infty d\tilde v \delta(\tilde v-\epsilon)e^{-\hat H_2S}\delta(\tilde v-\epsilon).
\ee

We expand the $\delta$-function in terms of the eigenfunctions as $\delta(\tilde v-\eps)=\sqrt{\frac{2}{\pi}}\int_0^\infty dq g(q)e^{-k\tilde v/2D}\sin(q\tilde v)$ where $g(q)=\sqrt{\frac{2}{\pi}}\int_0^\infty d\tilde v e^{k\tilde v/2D}\delta(\tilde v-\eps)\sin(q\tilde v)$. Applying this to the second delta function and Taylor expanding in the lowest order of $\eps$, we arrive at
\bea\label{eq:Z_e(S)}
Z_\eps(S)&=&\frac{2}{\pi}\int_0^{\infty}dq\sin^2(q\epsilon)e^{-D\lb\lb\frac{k}{2D}\rb^2+q^2\rb S}\nonumber\\
&\approx&\sqrt{\frac{2}{\pi}}\frac{\eps^2}{(2DS)^{3/2}}e^{-\frac{1}{4D}k^2S}.
\eea
Similarly, the numerator path integral in Eq.~(\ref{eq:Cvm_S}) is determined by expanding in the discrete set of eigenfunction of the Hamiltonian $\hat H_2$ with square-well boundary conditions at $\tilde v=0$ and $\tilde v=v_m$. The eigenfunctions are
\begin{equation}
f_n(\tilde v)=\sqrt{\frac{2}{v_m}}\sin^2\lb \frac{n\pi \tilde v}{v_m}\rb e^{-k\tilde v/2D},
\end{equation}
with the eigenvalues $E_n=D\left(\frac{n^2\pi^2}{v_m^2}+\lb\frac{k}{2D}\rb^2\right)$. Therefore, we obtain that 
\bea
C_\eps(v_m|S)&=&\frac{1}{Z_\eps(S)}\frac{2}{v_m}\sum_{i=1}^{\infty}\sin^2\lb\frac{n\pi\epsilon}{v_m}\rb \nonumber\\
&\times &\exp\left(-DS \lb \frac{n^2\pi^2}{v_m^2}+\lb\frac{k}{2D}\rb^2\rb\right),
\eea
which, in the limit of $\eps\rightarrow 0$, leads to 
\bea\label{cumulative}
C(v_m|S)
&=&\frac{\sqrt{2\pi}(2DS)^{3/2}}{v_m}\sum_{n=1}^\infty\lb\frac{n\pi}{v_m}\rb^2\nonumber\\
&\times&\exp\left(-\frac{\pi^2n^2DS}{v_m^2}\right).
\eea
Interestingly, $k$ has canceled completely out of the final answer.

Differentiating with respect to $v_m$ in Eq.~(\ref{cumulative}), we obtain the PDF for $v_m$ given by the scaling form 
\be\label{eq:sizeScalingForm}
P(v_m|S)=\frac{1}{\sqrt{2DS}}F_S\lb\frac{ v_m}{\sqrt{2 DS}}\rb,
\ee
where $F_S(x)$ is the Brownian excursion scaling function 
\be\label{eq:sizeScalingFunc}
F_S(x)=\frac{\sqrt{2\pi}}{x^4}\sum_{n=0}^\infty n^2\pi^2\lb\frac{n^2\pi^2}{x^2}-3\rb e^{-\frac{n^2\pi^2}{2x^2}}.
\ee
Using the Poisson summation formula \cite{Lighthill1958}
\be
\sum_{n=\infty}^\infty e^{-ny^2}=\sqrt{\frac{\pi}{y}}\sum_{n=\infty}^\infty e^{-n^2\pi^2/y^2},
\ee
the scaling function can be written
\be
F_S(x)=\sum_{n=1}^\infty(32n^4x^3-24n^2x)e^{-2n^2x^2},
\ee
and we can see it has the asymptotic form
\be
F_S(x)=
\left\{\begin{array}{ll}
	\sqrt{2\pi^5}\left(\frac{\pi^2}{x^{6}}-\frac{3}{x^{4}}\right)e^{-\frac{\pi^2}{2x^2}}&, x\rightarrow0\\ \\(32x^3-24x)e^{-2x^2}&, x\rightarrow\infty,
\end{array}\right.
\ee
which guarantees that all of its moments are finite. In Fig.~\ref{fig:3}, we present the numerically computed distribution of maximal velocities for different avalanche sizes which agrees very well with the analytical result. The rescaled PDF $P(v_m|S)$ collapses onto a scaling form as predicted by Eq.~(\ref{eq:sizeScalingFunc}).

\section{Overall duration, size and maximum velocity distributions}\label{sect:OverallDists}
%
\begin{figure}[!ht]
\includegraphics[trim=1cm 0 0 0,width=0.45\textwidth]{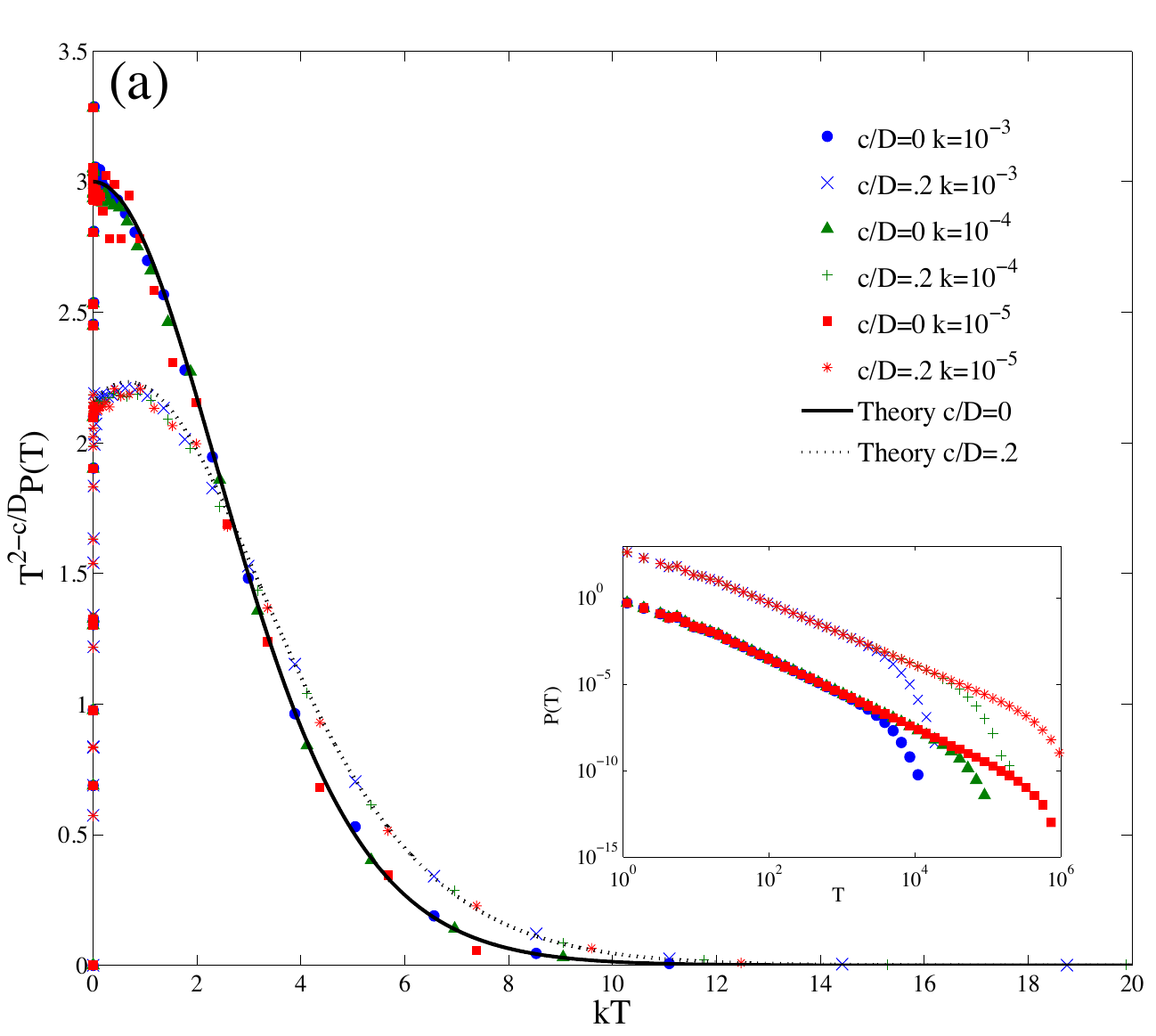}
\includegraphics[trim=1cm 0 0 0,width=0.45\textwidth]{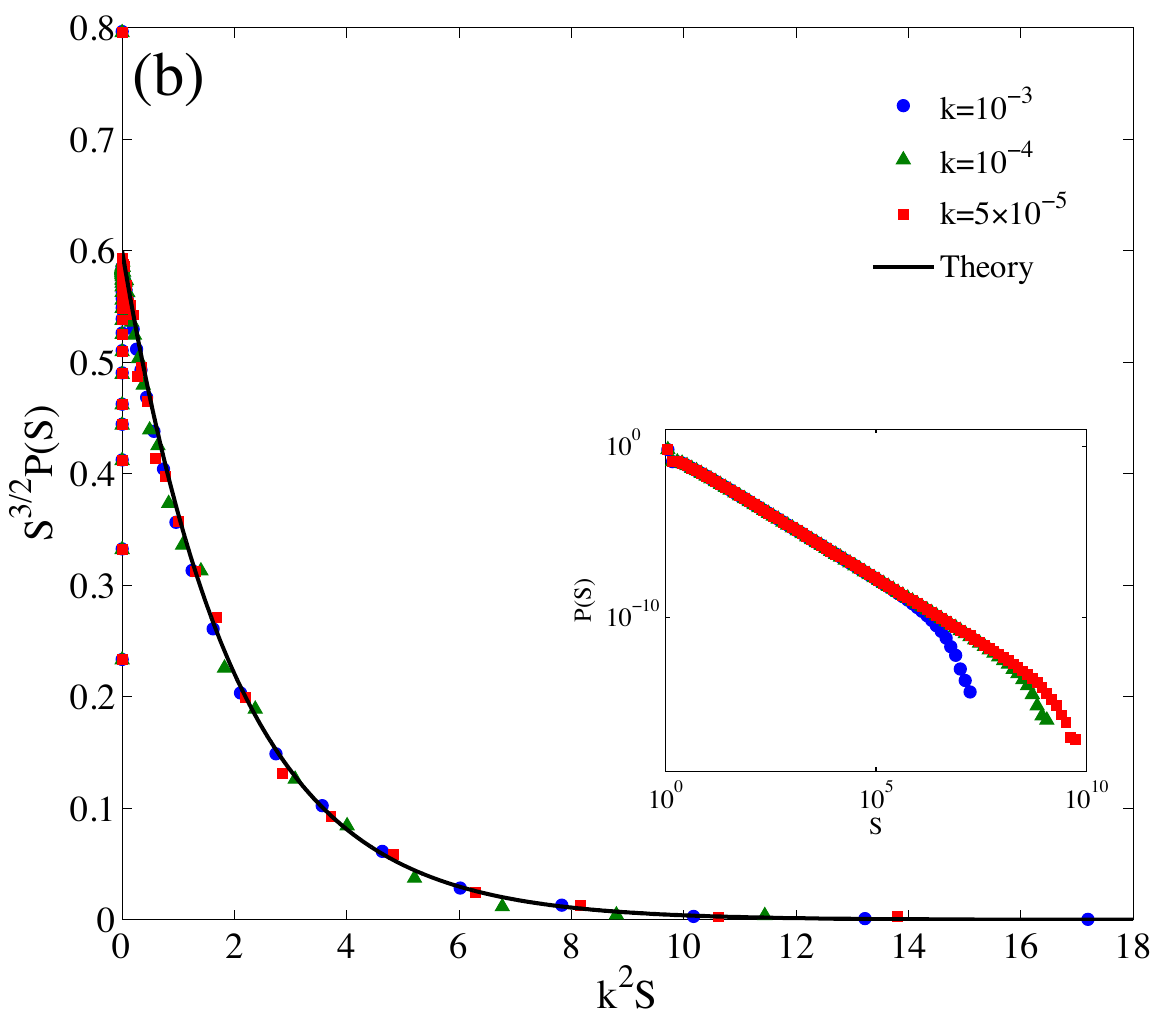}
\caption{ (Color online) (a) Numerical scaling collapse of the avalanche duration distribution $P(T)$ for different values of $k$ (see the caption to Table~\ref{table:tbl1} for the definition) and the driving rate $c\equiv\langle v\rangle/k$ to the form of Eq.~(\ref{eq:durDist2}). The scaling functions plotted are $F_{\tilde c}(x)=C(\tilde c)e^xx^{2-\tilde c}(e^x-1)^{\tilde c-2}$ which are proportional to the ones given in Eq.~(\ref{eq:durDist2}). The non-universal constants are fit by eye as $C(0)=3.0$ and $C(.2)=3.375.$ The inset shows the same distributions before rescaling with the curves for $\tilde c=0.2$ offset for visibility. (b) Collapse of the size distributions for different values of $k$ to the form of Eq.~(\ref{eq:sizeDist}). The theoretical curve plotted is $F(x)=C\exp(-x/2)$ where $C=0.6$ which is, up to normalization, what is given in Eq.~(\ref{eq:sizeDist}) with $D=1/2.$ The inset shows the size distributions before rescaling.} \label{fig:4}
\end{figure}

The overall duration distribution is proportional to the probability that an avalanche returns to the origin at time $T,$ assuming we use absorbing boundary conditions to guarantee that it will remain strictly positive in between. This means that, from Eq.~(\ref{eq:exactProp}), we have 
\bea\label{eq:durDist1}
P(T)&=&\lim_{\eps\rightarrow0}N(\eps)P(\eps;T,\eps,0)\nonumber\\&=&\lim_{\eps\rightarrow0}N(\eps)\lb\frac{2k}{1-e^{-kT}}\rb^{2-\tilde c}\frac{e^{(\tilde c-1)kT}\epsilon^{1-\tilde c}}{\Gamma(2-\tilde c)},
\eea
where $N(\eps)$ is a proportionality constant that cannot simply be set by normalization since $P(\eps,T;\eps,0)\sim T^{-2+\tilde c}$ for $T\rightarrow0$ and the normalization integral diverges at this limit. This divergence must occur because we expect that the return time to the origin gets shorter as the starting value $\eps$ gets closer to zero. To get rid of it, we need to set a cutoff $T^*\sim\eps$ defined as the minimum duration of an observable avalanche. The natural long duration cutoff is set by $1/k$. Assuming that these cutoffs are well-separated, namely that $kT^*\ll 1$, we have
\be\label{eq:durDist2}
P(T)=k(1-\tilde c)(kT^*)^{1-\tilde c}e^{kT}\lb\frac{1}{e^{kT}-1}\rb^{2-\tilde c}
\ee
for $T> T^*.$
This has power law behavior $P(T)\sim T^{-\alpha}\mathcal{G}(kT)$ with exponent $\alpha=2-\tilde c$ and $\mathcal{G}(x)$ a cutoff function, as can be seen by taking the limit $k\rightarrow0.$ The exponent agrees with the one reported in Refs.~\cite{Colaiori08,Zapperi98}. In Fig.~\ref{fig:4}a we numerically verify Eq.~(\ref{eq:durDist2}) for different values of $c$ and $k$.

The avalanche size distribution can be computed similarly. Starting from Eq.(\ref{eq:ABBM_vu}) with $k=0$ as in Ref.~\cite{Colaiori08} and using the results above for a random walk in a logarithmic potential, one can obtain the exponent $\tau=(3-\tilde c)/2.$ For $k\neq 0$ and $c=0,$ we can use Eq.~(\ref{eq:Z_e(S)}) and normalize with a cutoff $S^*\sim \eps^2$ as we did with the duration distribution, obtaining
\be\label{eq:sizeDist}
P(S)=\frac{\sqrt{S^*}}{2S^{3/2}}e^{-\frac{1}{4D}k^2S}.
\ee

The tail behavior of the overall distribution $P(v_m)$ of avalanche maxima can then be derived in the adiabatic limit by integrating Eq.~(\ref{eq:sizeScalingFunc}) against the overall size distribution Eq.~(\ref{eq:sizeDist}),
\bea
P(v_m)&=&\nonumber\int_{S^*}^\infty P(v_m|S)P(S)dS\\&=&\frac{\sqrt{2DS^*}}{2v_m^2}\int_0^{\frac{v_m}{\sqrt{2DS^*}}}xF_S(x)e^{-\frac{1}{2}\lb\frac{kv_m}{2Dx}\rb^2}dx.
\eea
The asymptotic behavior of the integrand allows us to take the upper limit of the integral to infinity provided $v_m\gg\sqrt{2DS^*}$. The infinite sum can then be integrated term by term using the formula
\be
\int_0^{\infty}x^ne^{-Ax^2-B/x^2} dx =A^{-\frac{n+1}{4}}B^{\frac{n+1}{4}}K_{\frac{n+1}{2}}\lb2\sqrt{AB}\rb,
\ee
where $K_\nu(x)$ is the modified Bessel function of the second kind. In this way, the correction factor to $v_m^{-2}$-scaling becomes 
\bea
\int_0^\infty xF_S(x)e^{-\frac{1}{2}\lb\frac{kv_m}{2Dx}\rb^2}dx\nonumber=\sqrt{\frac{\pi}{2}}\lb\frac{kv_m}{D}\rb^2\sum_{n=1}^\infty ne^{-\frac{nkv_m}{D}}.\\
\eea
The infinite series can be summed using the formula $\sum_n n \exp(- n x) = 4/\sinh^2(x/2)$, giving
\bea
\int_0^\infty xF_S(x)e^{-\frac{1}{2}\lb\frac{kv_m}{2Dx}\rb^2}dx=\sqrt{8\pi}\lb\frac{\frac{kv_m}{D}}{\sinh\lb \frac{kv_m}{2D}\rb}\rb^2.
\eea

Normalizing with a small velocity cutoff $v_m^*\sim\sqrt{2DS^*},$ the distribution is 
\be\label{eq:P(v_m)}
P(v_m)=\frac{v_m^*}{v_m^2}\lb\frac{\lb\frac{kv_m}{2D}\rb}{\sinh\lb\frac{kv_m}{2D}\rb}\rb^2,
\ee
which is valid for $v_m\gg v_m^*.$ We have already seen in Ref.~\cite{LeBlanc12} that the driving rate dependence of the power law $P(v_m)\sim v_m^{-\mu}$ at criticality is $\mu=2-\tilde c.$ In Fig.~\ref{fig:3}b, we confirm Eq.~(\ref{eq:P(v_m)}) numerically. 

Although we have derived the exact functional form of $P(v_m)$, it is important to note that the $\mu=-2$ exponent follows essentially from two facts. The first is that the scaling forms given by Eqs.~(\ref{eq:scalingForm}) and~(\ref{eq:sizeScalingForm}) have the exponents one would assume from dimensional analysis, and the second is that the integral leading to Eq.~(\ref{eq:P(v_m)}) converges when its bounds are taken to $0$ and $\infty$. Under these conditions, the fluctuations in $v_m$ are distributed narrowly about the average value $\langle v_m|T\rangle\sim T$, and so we expect the maximum velocity to have the same scaling exponent as the duration. However, there is no guarantee that these conditions hold in all cases.

In fact, this simple scaling by dimensional analysis runs contrary to what one might normally expect for an extreme value distribution. For example, if one wanted to make a simple argument, one could ignore correlations and boundary conditions and crudely approximate the avalanche as a sequence of $N=T/\delta t$ independent, exponentially distributed variables with average value $\langle v|T\rangle =AT$ where $\delta t$ is the duration of a time step and $A$ is a proportionality constant with the same dimensions as $D$. Then, the probability distribution function of the maximum value is given by $P(v_m|T) = \frac{N}{AT}e^{-v_m/AT}\left(1-e^{-v_m/AT}\right)^{N-1}$. The expectation value of $v_m$ can be determined by using the binomial expansion $P(v_m|T) = \frac{N}{AT}\sum_{k=0}^{N-1}\frac{(N-1)!(-1)^k}{k!(N-1-k)!}e^{-(k+1)v_m/AT}$. The integration corresponding to the first moment can be performed term by term, and the average maximum takes the form $\langle v_m|T\rangle =\frac{N}{AT} \sum_{k=0}^{N-1}\frac{(N-1)!(-1)^k}{k!(N-1-k)!} \frac{(AT)^2}{(k+1)^2}$. This is equivalent to $\langle v_m|T\rangle = AT\left(\Psi(N+1)+\gamma\right)$ where $\Psi(x)=\frac{d\log \Gamma(x)}{dx}$ and $\gamma$ is Euler's constant. As $T\rightarrow \infty$, this expression scales asymptotically as $AT\log (T/\delta t)$. The logarithmic enhancement is due to the fact that the random variable has $T/\delta t$ independent tries to achieve a value well above its average. Integrating this against the duration distribution $P(T)\sim T^{-2}$ gives a maximum value distribution with leading order behavior $P(v_m)\sim \log(v_m)v_m^{-2},$ a slower decay than the exact answer $P(v_m)\sim v_m^{-2}.$ However, temporal correlations change the picture considerably, preventing the average maximum value from scaling faster than the average, so the actual scaling law is $\langle v_m|T\rangle\sim\langle v|T\rangle \sim T$ which corresponds to $P(v_m)\sim v_m^{-2}.$


In Table~\ref{table:tbl1}, we summarize the quasistatic mean field theory results for the scaling functions and exponents, incorporating the new results for the maximum avalanche velocity. However, it should be stressed that these cutoff exponents have not been observed in all cases in experiments and simulations with long-ranged forces, even when the power law scaling exponents are mean field. In Ref.~\cite{Durin2000}, it is argued that in the  case when the cutoff is caused by a demagnetizing field coupled \emph{globally}, with $\partial_t u=-k\int u(t,x)d^2x,$ instead of the local coupling $\partial_tu=-ku(x,t),$ the cutoff exponents are different from the mean field predictions in a way that follows from dimensional analysis and is consistent with Barkhausen noise experiments and long-range simulations. In our case it would change the maximum velocity cutoff exponent from $\rho=1$ to $\rho_k=\frac{1}{3}.$ The effects cancel out and do not alter the exponent products that appear in the last three rows of Table~\ref{table:tbl1}. In other cases \cite{Zaiser02}, deviation of the cutoff exponents from the mean field values may be caused by hardening, an effect not accounted for by our considerations.

\section{Discussion and conclusions}
We have calculated exact scaling functions and exponents for the maximum avalanche velocity statistics in a mean field approximation of interface depinning. The distribution of maximum events in avalanches of fixed duration has a robust scaling form with a scaling function that depends on the driving rate, while the distribution for fixed avalanche sizes is independent of the elastic coupling constant in the adiabatic limit. The statistics of maximum velocities in arbitrary avalanches follows by integrating the maximum value distribution for fixed avalanche size or duration against the size or duration distribution. We find that the distribution of peak velocities in a train of non-overlapping avalanches has a scaling regime with power-law exponent $\mu =2-\tilde c$ followed by cut-off regime for $v_m>D/k$. Although the mean field theory captures very well the universal statistical properties of avalanches, i.e., the size, duration, and maximum value of plastic slip avalanches, it still remains an open problem as to what extent it can also describe other statistical quantities, such as velocity fluctuations during plastic deformations. For instance, discrete dislocation simulations report that the individual dislocation velocity probability distribution follows a power law with exponent $\sim -2.5$~\cite{miguel2001idf}, whereas Eq.~(\ref{eq:steadyState}) from mean field theory predicts an exponent of $-1$ for the collective velocity distribution.

Based on our calculations, we expect the $v_m^{-\mu}$ power law scaling of the maximum velocity distribution to be a very robust experimental observable, even in the case of poor time resolution where the true maximum might be missed. In fact, even if all an experiment could accomplish was to measure a random velocity within each avalanche, the inverse squared power law prediction does not change. This can be seen by a simple scaling argument. Namely, Eq.~(\ref{eq:scalingForm}) predicts that the average value of the maximum velocity scales linearly with the avalanche duration as $\langle v_m|T\rangle \sim T$, thus, in the same way as the average velocity $\langle v|T\rangle = \langle S|T \rangle/T\sim T.$ Since the maximum does not outpace the average as the avalanches get larger, we expect the same power-law behavior for the distribution regardless of whether we sample the maximum or a random point in the avalanche signal. This implies that even if the maxima are taken from a low-resolution time series, one should see an inverse squared power law distribution in the limit of slow driving, and this may make the maximum velocity easier to work with than the duration in low-resolution experiments. The scaling $\langle v_m|S\rangle\sim S^{1/2}$ should show similar independence from how well $v_m$ is able to be measured.

However, one thing that can affect the exponent substantially is the driving rate, which must be very slow in order to see $\mu=2$. In fact the variability in the acoustic emission experiments~\cite{Weiss97,Weiss2000,miguel2001idf,Weiss05,Weiss07,Fressengeas09} on the value of $\mu$ might  be related to the fact that the experiments are performed at nonzero driving rate. If that is the case, we find that the deviation from the adiabatic driving rate enters in the mean field exponents as $\mu =2-\tilde c$, in a way similar to that for the driving-rate dependent scaling exponents $\alpha$ and $\tau$ in the power-law distributions of avalanche durations and sizes. Experiments where the power law exponent is studied as a function of applied shear rate could test this.

Finally, we have shown that exact results for scaling functions in the maximum velocity distributions are easy to derive, even including nonzero sweep rates and tuning away from criticality. For instance, our results for the maximum velocity distributions $P(v_m|T)$ and $P(v_m|S)$ given duration and size, given in Eqs.~(\ref{eq:scalingForm}) and~(\ref{eq:sizeScalingForm}) respectively have simple scaling forms, the second of which is independent of the distance from criticality $k.$ Though it might be difficult to obtain sufficient statistics to check the scaling functions of these distributions experimentally, the predicted scaling of the conditional moments $\langle v_m^n|T\rangle\sim T^n$ and $\langle v_m^n|S\rangle\sim S^{n/2}$ could be checked much more easily. This could serve as a starting point for new precision tests of universality in the Barkhausen effect where high resolution time series are obtainable and there is already strong evidence of mean-field behavior \cite{Colaiori08}.

\noindent {\it Acknowledgments:} This work
was partially supported by the National Science Foundation through
grants DMS-1069224, and DMR-1005209. L.\,A. is grateful for support through a grant from the Center of Excellence for Physics of Geological Processes. M.L. and K.D. thank Braden Brinkman, Georgios Tsekenis, Nir Friedman, Matt Wraith, Stefano Zapperi and J.T. Uhl for useful discussions. 

\bibliographystyle{apsrev4-1}
\bibliography{references} 

\end{document}